\documentstyle[pre,aps]{revtex}
\begin{document}
\draft
\title{Statistical predictability in two-dimensional turbulence}

\author{G. Boffetta and S. Musacchio}
\address{Dipartimento di Fisica Generale, Universit\`a di Torino,
         Via Pietro Giuria 1, 10125 Torino, Italy}
\address{and INFM Sezione di Torino Universit\`a, Italy}

\date{\today}

\footnotetext{{\it Corresponding author address:} Guido Boffetta,
Dipartimento di Fisica Generale, via P. Giuria 1, 10125 Torino, Italy.
Email: boffetta@to.infn.it}                                                     

\maketitle
%\newpage

\begin{abstract}
The predictability problem in the inverse energy cascade of
two-dimensional turbulence is addressed by means of high
resolution direct numerical simulations. 
The analysis is done in terms of the finite size Lyapunov exponent
(FSLE) which is a measure of the growth rate at fixed error
level.
The numerical results are compared with classical closure 
predictions and good quantitative agreement is found.
Finally, it is shown that the inertial range predictability
properties are not affected by the presence of noise induced
by the small, not resolved, scales.
\end{abstract}

%\newpage
%%%%%%%%%%%%%%%%%%%%%%%%%%%%%%%%%%%%%%%%%%%%%%%%%%%%%%%%%%%%%%
\section{Introduction}
\label{sec:1}
The knowledge of predictability properties of two--dimensional
turbulence is fundamental for estimating the predictability of 
atmospheric flow 
%\cite{Leith71,LK72}. 
(Leith 1971, Leith and Kraichnan 1972).
The characteristic property of two-dimensional flow is enstrophy 
conservation which force the energy to flow toward large scales. 
In this inverse cascade regime, dimensional arguments predict a 
``5/3'' Kolmogorov energy spectrum which is indeed observed in the 
large-scale atmospheric spectrum
%\cite{NGJ84}.
(Nastrom, Gage and Jasperson 1984).

From a theoretical point of view, predictability in fully
developed turbulence has been investigated as a prototypical
model with many characteristic scales and time.
The first attempts to the study of predictability in turbulence
dates back to the pioneering work of Lorenz
%\cite{Lorenz69}
(Lorenz 1969)
and to Kraichnan and Leith papers 
%\cite{Leith71,LK72}.
(Leith 1971, Leith and Kraichnan 1972).
On the basis of closure approximations, they were able to obtain
quantitative predictions on the evolution of the error in different
turbulent situations, both in two and three dimensions. 
Their fundamental papers become the backbone for more recent
approaches
%\cite{Lesieur97}.
(Lesieur 1997).
Because predictability experiments in fully developed turbulence
are numerically very expensive, to our knowledge there are still no
attempt to compare closure results with direct numerical simulations.

In this paper we address the predictability problem for 
inverse energy cascade of two--dimensional turbulence by means 
of high resolution direct numerical simulations.
It has been recently shown that inverse cascade is
not affected by intermittency corrections and velocity
statistics is quasi--Gaussian
%\cite{PT97,BCV00}.
(Paret and Tabeling 1997, Boffetta, Celani and Vergassola 2000).
This makes the problem simpler than the three--dimensional case
and we expect that closure-based predictions are essentially correct.

At variance with direct cascade, the inverse energy cascade
cannot be observed in decaying turbulence. In order to 
sustain the cascade, a continuous input of energy by random
forcing at small scales is necessary. 
Because we are interested in the intrinsic predictability of the
model, we will study the problem assuming known the realization
of the forcing.
In realistic applications, the forcing represents the small scale 
dynamics not resolved by the two--dimensional model (i.e. convective 
motion in the atmosphere). In this case one should take into
account also the uncertainty introduced by the random forcing.
Because of the hierarchical structure of the characteristic times,
we will see that the inertial range predictability properties are 
not affected by the uncertainty introduced by the forcing.

This remainder of the paper is organized as follows: Section
\ref{sec:2} is devoted to a brief summary of the classical closure
results. In Section \ref{sec:3} and we present the numerical results
and their comparison with predictions.
Section \ref{sec:4} is devoted to conclusions.

%%%%%%%%%%%%%%%%%%%%%%%%%%%%%%%%%%%%%%%%%%%%%%%%%%%%%%%%%%%%%%%%
\section{Statistical turbulence predictability}
\label{sec:2}
Given two realizations of the velocity field
${\bf u}^{(1)}({\bf x},t)$ and ${\bf u}^{(2)}({\bf x},t)$, a suitable
measure for the predictability is the error field
\begin{equation}
\delta {\bf u}({\bf x},t) = {1 \over \sqrt{2}} \left(
{\bf u}^{(2)}({\bf x},t) - {\bf u}^{(1)}({\bf x},t) \right)
\label{eq:2.1}
\end{equation}
from which we define the error energy and the error energy spectrum as
%\cite{LK72,Lesieur97}
(Leith and Kraichnan 1972, Lesieur 1997)
\begin{equation}
E_{\Delta}(t) = {1 \over 2} \int | \delta {\bf u}({\bf x},t) |^2 d^2 x
= \int_{0}^{\infty} E_{\Delta}(k,t) dk \, .
\label{eq:2.2}
\end{equation}
Normalization in (\ref{eq:2.1}) ensures that $E_{\Delta}(k,t) \to E(k)$
for uncorrelated fields (i.e. for $t \to \infty$).

The two realizations of the turbulent flow are stationary solutions
of two--dimensional Navier Stokes equation in which the inverse
cascade is sustained by energy injection at small-scale 
wavenumber $k_{f}$ with a Gaussian random forcing.
As it is customary in predictability experiments, 
for most of the results presented here the two realizations of the
forcing will be the same. In this case the growth of the error will
be uniquely due to the deterministic chaotic dynamics. 
At the end of Section \ref{sec:3} we will discuss the case 
in which two different realizations of the forcing are implemented.

Assuming that the initial error can be considered infinitesimal,
the magnitude of the difference field starts growing exponentially and
$E_{\Delta}(t) \simeq E_{\Delta}(0) \exp(L(2) t)$
where $L(2)$ is the generalized Lyapunov exponent (in non--intermittent
systems, $L(2)=2 \lambda$, see 
%\cite{BJPV98}
(Bohr, Jensen, Paladin and Vulpiani 1998).
The error growth in this stage is confined at the
faster scales in the inertial range, corresponding in our situation
to the scales close to the forcing wavenumber $k_{f}$, while at larger scales
the two fields remain correlated (see Figure~\ref{fig:1}).
At larger times, when $E_{\Delta}(k_{f},t)$ becomes
comparable with $E(k_{f})$, the exponential growth terminates, because the
two fields are completely decorrelated at small scales.
The error growth continues at larger scales in the inertial range,
where the two fields are still correlated, and the time evolution of the 
error follows an algebraic law.

The error transfer towards large scale can be quantified by the
characteristic wavenumber $k_{E}(t)$ at which the error spectrum
is a given fraction of the reference spectrum.
Leith and Kraichnan
%\cite{LK72}
(Leith and Kraichnan 1972),
defining $k_{E}$ such that the relative error spectrum 
$r(k,t) = E_{\Delta}(k,t)/E(k)$ is $r(k_E)=0.5$, 
found a self-similar error spectrum $r(k/k_E)$ in which the time 
dependence is only through $k_E(t)$.

The dimensional prediction, based on the assumption that the time
it takes for the error to induce a complete uncertainty at wavenumber
$k$ is proportional to the characteristic time at that scale gives,
within the Kolmogorov scaling
\begin{equation}
{d k_{E} \over d t} = B \varepsilon^{1/3} k_{E}^{-5/3}
\label{eq:2.3}
\end{equation}
where $B$ is an adimensional constant. By integration one gets
\begin{equation}
 k_{E}(t)^{-2/3} = k_{0}^{-2/3} + B \varepsilon^{1/3} (t-t_0)
\label{eq:2.4}
\end{equation}
If the inertial range is sufficiently wide, we can assume that
$k_0 \ll k_E$ and thus $k_E(t) \sim \varepsilon^{-1/2} t^{-3/2}$
%\cite{Lorenz69}. 
(Lorenz 1969).
At wavenumbers smaller than $k_E$ the error is still small in comparison
with the typical energy, while at larger wavenumbers the two fields are
completely decorrelated. Thus one can assume that in (\ref{eq:2.2}) one
has
\begin{equation}
E_{\Delta}(k) = \left\{
\begin{array}{ll}
E(k) & \mbox{for} k>k_E(t) \\
0 & \mbox{for} k<k_E(t)
\end{array}
\right.
\label{eq:2.5}
\end{equation}
Using the Kolmogorov spectrum for $E(k)$ one ends with the prediction 
\begin{equation}
E_{\Delta}(t) = G \varepsilon t
\label{eq:2.6}
\end{equation}
The numerical constant $G$ in (\ref{eq:2.6}) can be obtained only
by repeating the argument more formally within a closure
framework. In the case of two--dimensional inverse cascade 
the Test Field model predicts $G \simeq 4.19$
(Leith and Kraichnan 1972).

Under the hypothesis of an initial error localized on infinitesimal scale in  
a wide inertial range (i.e. $k_(0) \to \infty$), reverting equation 
(\ref{eq:2.4}) is possible to relate 
the predictability time on a certain scale with the 
characteristic time taken from the energy spectrum:
\begin{equation} 
T_p(k) = {1 \over B} \varepsilon^{-1/3} k^{-2/3} = {C^{1/2} \over B}
{\left[ E(k) k^3 \right]}^{-1/2} \equiv {C^{1/2} \over B} \tau(k)
\label{eq:2.7}
\end{equation}
where $C$ is Kolmogorov constant ($C \simeq 6$ in two dimension)
and $\tau(k)$ is the spectrum--base characteristic time.
This result has a practical utility, because in atmospheric turbulence 
is obviously impossible to perform a ``predictability experiment'' and
equation (\ref{eq:2.7}) give an estimate for predictability time  
simply measuring the energy spectrum. 

An alternative approach to predictability problem is the 
recently introduced Finite Size Lyapunov Exponent analysis 
%\cite{ABCPV96}
(Aurell et al. 1996).
FSLE is a generalization of the Lyapunov exponent to
finite size errors. Within this approach one computes the 
error doubling time $T_{r}(\delta)$ which takes for an error of size 
$\delta$ (with a given norm) to grow of a factor $r$ (typically $r=2$)
From the average doubling time one defines the FSLE as
\begin{equation}
\lambda(\delta)={1\over \langle T_{r}(\delta) \rangle}
\ln r
\label{eq:2.8}
\end{equation}
It is easy to show that definition (\ref{eq:2.8}) reduces
to the standard Lyapunov exponent $\lambda$ in the limit of infinitesimal
errors $\delta \to 0$
%\cite{ABCPV96}
(Aurell et al. 1996).
As the Lyapunov exponent can be seen like the inverse of the fastest 
characteristic time of a dynamical system, the FSLE can be considered  
the inverse of the characteristic time at a certain scale.  
For finite errors, the FSLE measures the effective
error growth rate at error size $\delta$.
Let us remark that taking averages at fixed time, as in (\ref{eq:2.6})
is not the same of averaging at fixed error size, as in (\ref{eq:2.8}).
This is particularly true in the case of intermittent systems,
in which strong fluctuations of the error in different realizations
can hide scaling laws in time.
From a numerical point of view, the computation of $\lambda(\delta)$
is not more expensive than the computation of the Lyapunov
exponent with a standard algorithm.   

In turbulence predictability, a natural measure of the error is 
$\delta= \sqrt{2E_{\Delta}}$.
The assumption of locality, i.e. that the FSLE is proportional to the
inverse of the characteristic time at the scale $k$ such that the
typical velocity is $u(k) \sim \delta$, 
combined with Kolmogorov scaling for the velocities
produces the scaling law for the FSLE
\begin{equation}
\lambda(\delta)=A \varepsilon \delta^{-2}
\label{eq:2.9}
\end{equation} 
The constant $A$ relates the energy flux in the cascade to the rate of 
error growth. Its value is not determined by dimensional arguments, but 
is easy to show that in absence of intermittency and for $r \simeq 1$  
it can be related to the constant $G$ in (\ref{eq:2.6}) by 
$A=ln(r^2)/(r^2-1)G$. In the limit $r \to 1$ one gets $A \to G$. 

Because of the appearance of the energy dissipation at the first power
in (\ref{eq:2.9}), we expect that this scaling law is universal,
i.e. not affected by possible intermittency in the velocity
statistics 
(Aurell et al. 1996).
The scaling law (\ref{eq:2.9}) is valid within 
the inertial range  $u(k_{f}) < \delta < U$ where $u(k_f)$ 
represents the typical
velocity fluctuation at forcing wavenumber and $U\simeq \sqrt{2 E}$
is the large scale velocity.
At large errors $\delta \simeq U$, we expect error saturation,
$E_{\Delta} \to E$ and thus $\lambda(\delta) \to 0$.    

%%%%%%%%%%%%%%%%%%%%%%%%%%%%%%%%%%%%%%%%%%%%%%%%%%%%%%%%%%%%%%%%
\section{Numerical simulations and analysis}
\label{sec:3}
We have performed extensive direct numerical simulation of the 
two--dimensional
Navier--Stokes equation written for the scalar vorticity 
$\omega({\bf x},t)=\nabla \times {\bf u}({\bf x},t)=-\triangle \psi({\bf x},t)$
as
\begin{equation}
\partial_t \omega + J(\omega,\psi) =
\nu \triangle \omega - \alpha \omega + f
\label{eq:3.1}
\end{equation}
where $J$ represents the Jacobian with the stream function $\psi$.
The friction term in (\ref{eq:3.1}) removes energy at large scales: 
it is necessary in order to avoid Bose--Einstein condensation on the 
gravest mode and to obtain a stationary state
%\cite
(Smith and Yakhot 1993). 
Physically this term represents the effect of bottom friction.
The random forcing $f$ is $\delta$--correlated in time and 
injects energy at wavenumber $k_{f}$ only.

Numerical integration of (\ref{eq:3.1}) is performed
by a standard pseudo-spectral code fully dealiased with second--order
Runge--Kutta time stepping on a doubly periodic square domain
with resolution $N=1024$.
As it is customary in numerical simulations, we use hyperviscous 
dissipation in order to extend the inertial range.

Stationary turbulent flow is obtained after a very long simulation
starting from a zero initial vorticity field. At stationarity one observes
a wide inertial range with a well developed Kolmogorov 
energy spectrum $E(k) = C \varepsilon^{2/3} k^{-5/3}$ (see Figure~\ref{fig:1}).
Structure functions in physical space are found in agreement 
with the self-similar Kolmogorov theory and no intermittency 
is detected
%\cite{BCV00}.
(Boffetta, Celani and Vergassola 2000).

The perturbed field is obtained from a configuration of the velocity field
${\bf u}^{(1)}({\bf r},0)$ in the stationary state,
\begin{equation}
{\bf u}^{(2)}({\bf r},0)={\bf u}^{(1)}({\bf r},0) +
\sqrt{2} \, \delta {\bf u}({\bf r},0)
\label{eq:3.9}
\end{equation}
in which the initial error $\delta {\bf u}({\bf r},0)$ is very small 
($E_{\Delta}(t=0) \simeq 10^{-5} E$), and 
is localized on small scales. The error energy spectrum (\ref{eq:2.1}) 
is thus initially localized at wavenumbers greater than the forcing 
wavenumber. In the inertial range the two fields are completely 
correlated at $t=0$.     
Let us remark that the precise form of the initial error spectrum
is not important provided that it can be considered infinitesimal,
at it will immediately evolve toward the direction of the first 
Lyapunov eigenvector.

The two configuration are integrated in time according to
(\ref{eq:3.1}) and the evolution of the error $\delta {\bf u}({\bf r},t)$
is computed. In this section we present the results obtained using
the same realization of random forcing in both the realizations.
In this way the growth of the error is only induced by the turbulent dynamic. 
At the end of this section we will compare this results with the 
ones obtained with two independent realizations of the forcing.

In Figure~\ref{fig:2} we plot the time evolution of the error energy 
$\langle E_{\Delta}(t) \rangle$ obtained from direct 
numerical simulations averaged over $20$ realizations.
The exponential regime is clearly visible at small times, showing that 
for infinitesimal error turbulence behaves exactly as a standard chaotic 
system.
At the very beginning is possible to observe an initial 
recorrelation of the two fields, i.e. a decreasing of $E_{\Delta}(t)$:
it takes a small, but finite time for the initial perturbation to 
align in the direction of the  
leading Lyapunov exponent. During this time the forcing and 
dissipation recorrelate the two fields.

Figure~\ref{fig:3} shows the computation of the FSLE from our simulations.
For very small errors, $\delta < u(k_f)$ corresponding to an error
spectrum $E_{\Delta}(k_f,t) << E(k_f)$, we observe the convergence of 
$\lambda(\delta)$ to the leading Lyapunov exponent 
$\lambda \simeq 1.07$. Its value is
essentially the inverse of the smallest characteristic time in the 
system and represents the growth rate of the most unstable features.
At larger errors $\delta > 10^{-2}$, we clearly see the transition to the
inertial range scaling (\ref{eq:2.9}). At further larger errors
$\delta \simeq U \simeq 0.1$,
$\lambda(\delta)$ falls down to zero in correspondence of error
saturation.

In order to emphasize scaling (\ref{eq:2.9}), in 
Figure~\ref{fig:3}
we also show the compensation of $\lambda(\delta)$ with 
$\varepsilon \delta^{-2}$. Prediction (\ref{eq:2.9}) is verified with
very high accuracy which allows to determine the value of 
$A = 3.9 \pm 0.1$. 
With the present value of $r \simeq 1.057$, this corresponds
to a value $G = 4.1 \pm 0.1$.
The physical picture we obtain is that the creation of
uncorrelated energy in the inertial range due to chaotic 
dynamics is about $4$ times faster than the energy transfer rate.
Our numerical result is in remarkable agreement with
the old prediction obtained within the
Test Field Model closure which gives $G=4.19$
%\cite
(Leith and Kraichnan 1972).

The physical meaning of $G$ is the ratio of the rate of
uncorrelated energy production to the rate of energy
injected by the forcing and transfered to large scales $\varepsilon$.
The fact that $G>1$ shows that the uncorrelated energy is not simply 
``transported'' with the energy through the cascade, 
but there is an effective production of error at each scale,
due to the chaotic dynamic. The constant rate of error--energy growth 
is not an immediate consequence of the constant energy flux 
in the inverse cascade, but is the effect of the dimensional
hypothesis that the times for energy transfer and error growing 
at a fixed scale should follow the same scaling law.
Moreover $G>1$ suggests that (\ref{eq:2.6}) should be 
unchanged when one considers two independent realizations of 
the forcing, as we will see below.

The time evolution of the error energy spectrum (Figure~\ref{fig:1})
shows a sharp front of error that propagates from small to large scales
justifying {\it a posteriori} the assumption (\ref{eq:2.5}).
From the error spectra in Figure~\ref{fig:1} one can compute the
characteristic wavenumber $k_E(t)$ defined as $r(k_E)=0.5$.
On scales smaller than $k_E(t)^{-1}$ the two fields are 
already decorrelated, while on larger scales the error can 
still be considered small, so we can define implicitly the 
predictability time $T_p(k)$ at certain wavenumber $k$ from
$k_{E}(T_p)=k$.

Figure~\ref{fig:4} shows the evolution of $k_E(t)^{-2/3}$ compared
with the best fit based on (\ref{eq:2.4}). The result is rather
noisy because of the discrete character of the wavenumbers, 
nevertheless it is possible to estimate the value of the constant
in (\ref{eq:2.4}) $B=0.43 \pm 0.02$.
In equation (\ref{eq:2.7}) $B$ relates the predictability time
to the characteristic time based on the spectrum. From our 
simulation we have $T_{p}(k) \simeq 5.7 \tau(k)$.

In order to check the prediction of the self-similar error growth we rescale
all the error spectra with the energy spectrum, 
obtaining the relative error spectra $r(k,t) = E_{\Delta}(k,t)/E(k)$.
Rescaling the wavenumbers with the characteristic wavenumber $k_E(t)$ the
error spectra collapse in a similarity error spectrum $r(k/k_E)$
(Figure~\ref{fig:5}).   
This means that in order to characterize statistically the 
error growth one needs to know only the evolution of $k_E(t)$. 
Closure computations also predict that the shape of the error spectrum
in the limit $k \to 0$ is $E_{\Delta}(k,t) \sim k^{-3}$ that for the 
similarity error spectrum gives $r(x) \sim x^{14/3}$ which is
indeed observed in our simulations.

\subsection{Different forcing}

All the results presented above are obtained forcing identically 
the two realizations of the turbulent field. 
In principle this is not consistent with a realistic application
like modeling large scale atmospheric flow because in that 
case the forcing is due to the motion on small scales that are non resolved.
In order to simulate this continuous injection of error it is more correct 
to use two independent realizations of the random forcing.
To check in which way this can affect the results discussed above, we have 
repeated the same numerical experiment with two independent forcing.

The main difference is that in this case the initial exponential growth
of the infinitesimal error is non present. When the scales smaller than 
the forcing scale are non completely decorrelated the main source of error
is due to the different forcing, not to the chaotic dynamic. 
If there were no dynamical effects 
we could expect to observe a ``diffusive'' behavior
and the error energy should grow linearly as in (\ref{eq:2.6}) but
because the source of the error is the energy input one should have
$G \simeq 1$.
This regime is indeed observed (Figure~\ref{fig:3}) but is worthwhile 
to remember that it is physically of little interest: we are using 
different forcing to mimic the unresolved small scale motion and so 
it is natural to assume that the two fields are completely 
decorrelated at those scales.

When this condition is satisfied, and the error is localized in the inertial 
range, it is not possible to distinguish the case with a different 
forcing from the one with one forcing. From the analysis of the FSLE    
(Figure~\ref{fig:3}) it is evident that in this range the scaling law 
(\ref{eq:2.9})
is verified with the same value of the constant $A$. 
All the inertial range results about 
the self similar growth of error are recovered.  
This demonstrate that the continuous injection of error due to an 
unresolved small--scale motion do not produce effects when the two 
turbulent fields are already decorrelated on such scales. 
This is a consequence of locality:
the error growth in the inertial range is completely determined by the 
non linear dynamic, and it is not sensible on what happens on smaller 
scale.      
                 
\section{Conclusion}
\label{sec:4}
We have studied the predictability problem in the inverse cascade of 
two--dimensional turbulence. 
Using the Finite Size Lyapunov Exponent we have shown that after an 
initial exponential growth, a linear growth of the decorrelated energy 
sets in. This decorrelated energy flows to large scales in self similar
way through an inverse cascade with a constant flux that is about four 
times faster than the energy flux. 

This behavior do not change also adding other possible perturbation 
on small scale, thus the results obtained in a numerical experiment with 
initial condition uncertainty can be applied to a wider class of problems.    
As an example the predictability time of mesoscale in the atmosphere 
can be estimated easily with a simple measure of the energy spectrum. 

All the results presented are in strong agreement with the prediction
of closure. 
At least from the point of view of predictability, two-dimensional
turbulence thus seems to be very well captured by low-order
closure scheme. 
As a consequence we can exclude, on the basis of our 
numerical findings, the existence of intermittency
effects in the inverse cascade of error.
This is a result which is probably of more 
general interest than the specific problem discussed in this article.

\section*{Acknowledgments}
We acknowledge the allocation of computer resources
from INFM Progetto Calcolo Parallelo.

%%%%%%%%%%%%%%%%%%%%%%%%%%%%%%%%%%%%%%%%%%%%%%%%%%%%%%%%%%%%%%%%%%%

%%%%%%%%%%%%%%%%%% FIGURE %%%%%%%%%%%%%%%%%%%%%%%%%%%%%%%%%%%%%%%
%\narrowtext
\begin{figure}
\caption{Stationary energy spectrum $E(k)$ (thick line) and 
error spectrum $E_{\Delta}(k,t)$ at time
$t=4.6, 5.5, 7.1, 10.0, 15.6$. $k_f=320$
is the forcing wavenumber. In the inset we plot the
compensated spectrum $\varepsilon^{-2/3}k^{5/3}E(k)$.}
\label{fig:1}
\end{figure}

\begin{figure}
\caption{Error energy $\langle E_{\Delta}(t) \rangle$ growth
averaged over $18$ runs.
Dashed line represents closure prediction (\ref{eq:2.6}), 
dotted line is the saturation value $E$. 
The initial exponential growth is emphasized by the lin-log
plot in the inset where the initial decreasing of the error is
also observable.}
\label{fig:2}
\end{figure}

\begin{figure}
\caption{Finite size Lyapunov exponent $\lambda(\delta)$ as a function
of velocity uncertainty $\delta$.
In the simulation with identical forcing for
the two fields ($+$) the asymptotic value for $\delta \to 0$ gives
the leading Lyapunov exponent of the turbulent flow. 
In the case with different forcing ($\times$) the infinitesimal regime
is unphysical. In the inertial range, $\delta \ge 10^{-2}$, the behavior 
of $\lambda(\delta)$ is identical for both case.
Dashed line represent the prediction (\ref{eq:2.9}).
In the inset we show the compensated plot
$\lambda(\delta) \delta^{2}/\varepsilon$. The line represent the
fit to the constant $A \simeq 3.9$.}
\label{fig:3}
\end{figure}

\begin{figure}
\caption{Time evolution of the characteristic wavenumber $k_E(t)^{-2/3}$
compared with the dimensional prediction
$k_E(t)^{-2/3} =k_0^{-2/3}+ B \varepsilon^{1/3}(t-t_0)$
where $t_0 = 6.0$ is the time that the error takes
to reach the inertial range at $k_0 = 133.0$}
\label{fig:4}
\end{figure}

\begin{figure}
\caption{Similarity error spectrum $r(k/k_E) = E_{\Delta}(k/k_E,t)/E(k/k_E)$.
All the error spectra rescaled collapse together within the error bands.
The log-log inset shows the asymptotic behavior $r(x) \sim x^{14/3}$ for
$x \to 0$}  
\label{fig:5}
\end{figure}

%%%%%%%%%%%%%%%%%%%%%%%%%%

\begin{references}

\bibitem[]{ABCPV96}
Aurell, E., G. Boffetta, A. Crisanti, G. Paladin, and A. Vulpiani,
1996:
Growth of non-infinitesimal perturbations in turbulence.
{\it Phys. Rev. Lett.}, {\bf 77}, 1262-1265.
 
\bibitem[]{ABCPV97}
Aurell, E., G. Boffetta, A. Crisanti, G. Paladin, and A. Vulpiani,
1997:
Predictability in the large: An extension of the concept of
Lyapunov exponent.
{\it J. Phys.}, {\bf A30}, 1--26.
 
\bibitem[]{BCV00}
Boffetta, G., A. Celani, and M. Vergassola, 2000:
Inverse energy cascade in two-dimensional turbulence: Deviations from
Gaussian behavior.
{\it Phys. Rev. E}, {\bf 61}, R29--32

\bibitem[]{BJPV98}
Bohr, T., M. Jensen, G. Paladin, and A. Vulpiani, 1998:
Dynamical Systems Approach to Turbulence,
Cambridge University Press, Cambridge, UK.

\bibitem[]{Leith71}
Leith, C. E., 1971:
Atmospheric predictability and two-dimensional turbulence.
{\it J. Atmos. Sci.}, {\bf 28}, 145--161.
 
\bibitem[]{LK72}
Leith, C. E., and R. H. Kraichnan, 1972:
Predictability of turbulent flows.
{\it J. Atmos. Sci.}, {\bf 29}, 1041--1058.
 
\bibitem[]{Lesieur97}
Lesieur, M., 1997:
Turbulence in Fluids, 3th edition, Kluver, Dordrecht.

\bibitem[]{Lorenz69}
Lorenz, E. N., 1969:
The predictability of a flow which possesses many scales of motion.
{\it Tellus}, {\bf 21}, 289--307.
 
\bibitem[]{NGJ84}
Nastrom, G.D., K.S. Gage, and W.H. Jasperson, 1984:
Kinetic energy spectrum of large and mesoscale atmospheric processes.
{\it Nature}, {\bf 310}, 36--38.

\bibitem[]{PT97}
Paret, J., and P. Tabeling, 1997:
Experimental Observation of the Two-Dimensional Inverse Energy
Cascade. {\it Phys. Rev. Lett.}, {\bf 79}, 4162--4165.

\bibitem[]{SY93}
Smith, L.M., and V. Yakhot, 1993:
Bose Condensation and Small-Scale Structure Generation in a
Random Force Driven 2D Turbulence.
{\it Phys. Rev. Lett.}, {\bf 71}, 352--355.

\end{references}
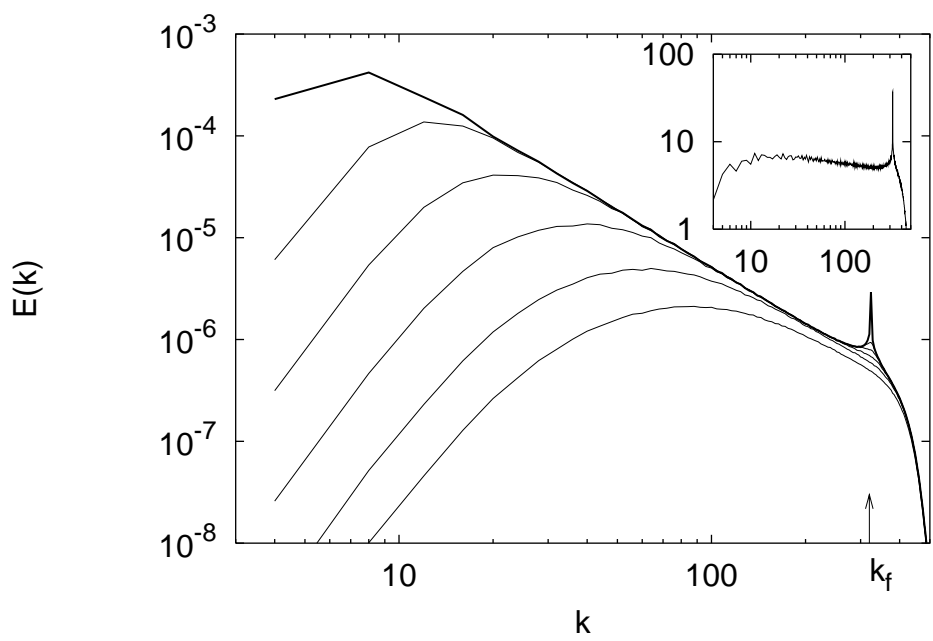
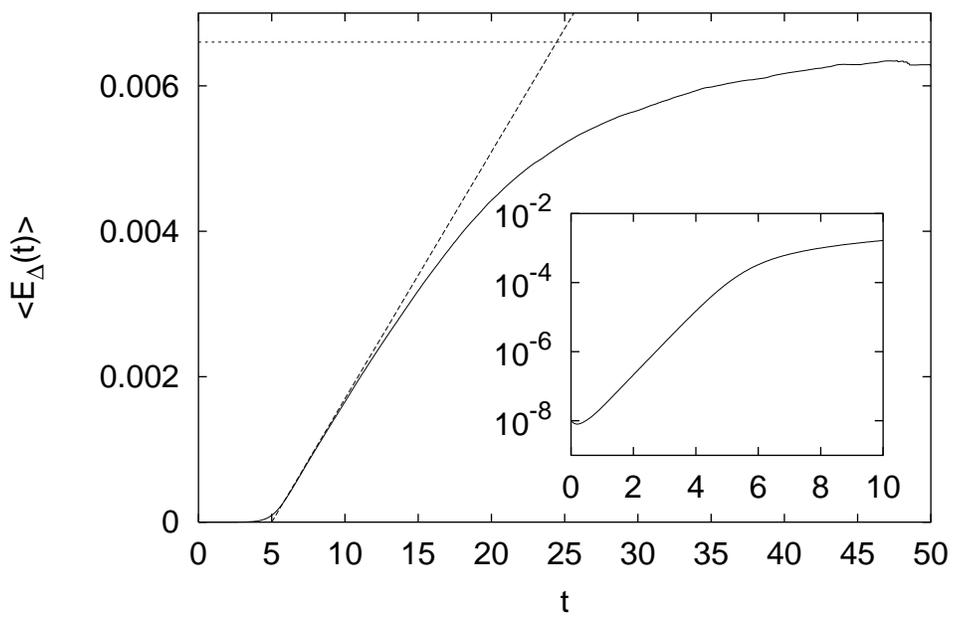
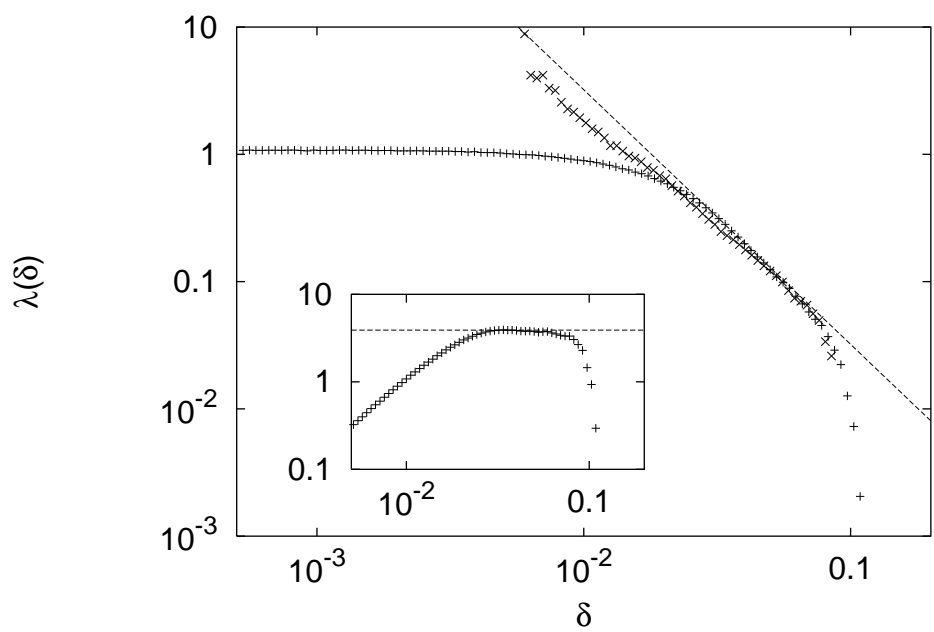
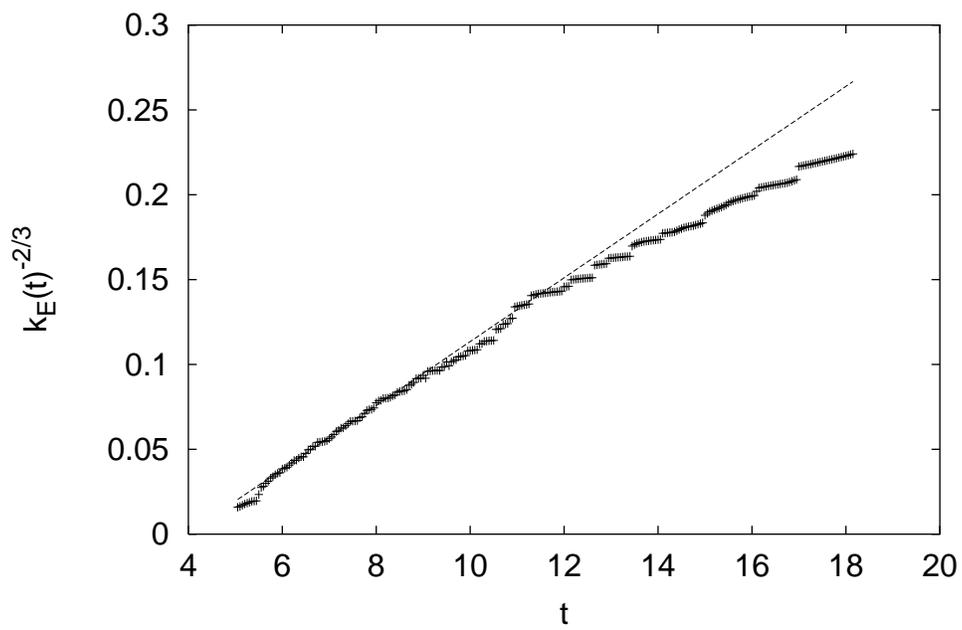
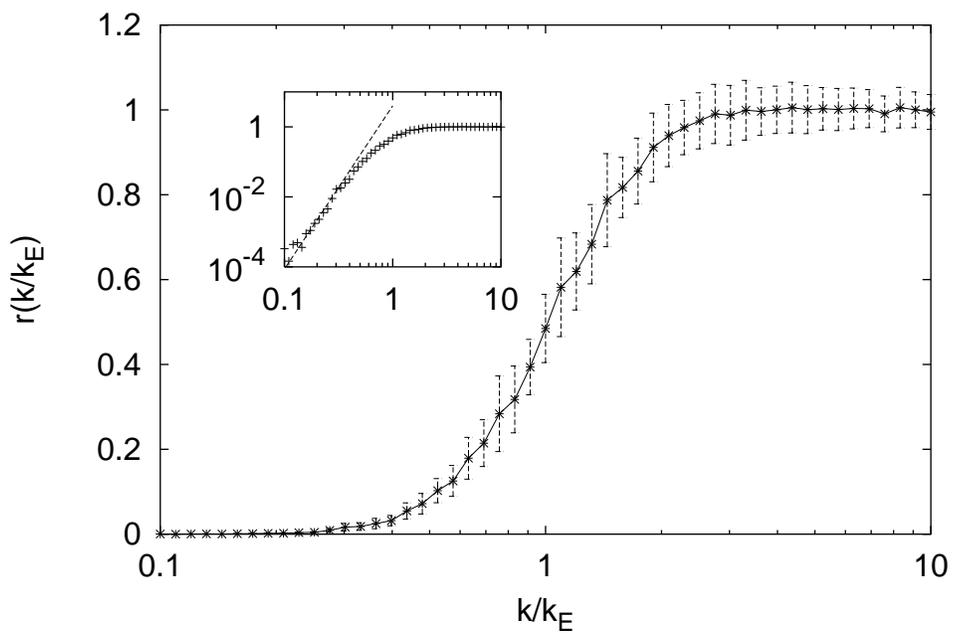
\end{document}